\documentclass[10pt,a4paper]{article}
\usepackage{amsmath}
\usepackage{amssymb}
\usepackage{amsfonts}
\usepackage{amstext}
\usepackage{amsbsy}
\usepackage[mathscr]{eucal}
\usepackage{graphicx}
\usepackage{color}
\usepackage[all]{xy}
\newcommand{\beq}{\begin{equation}}
\newcommand{\eeq}{\end{equation}}
\newcommand{\beqa}{\begin{eqnarray}}
\newcommand{\eeqa}{\end{eqnarray}}
\newcommand{\ket}[1]{| #1 \rangle}

\makeindex
\title{\Large\textbf{A class of quantum gate entangler}}

\author{\textit{ Hoshang Heydari}\\
        \small\textit{Physics Department, Stockholm university 10691 Stockholm Sweden}\\
\\\small\textit{Email: hoshang@fysik.su.se}}

\date{}
\begin{document}

\maketitle \thispagestyle{empty}

\maketitle
\begin{abstract}
We construct quantum gate entanglers for different classes of multipartite states based on  definition of W and GHZ concurrence classes. First, we review the basic construction of concurrence classes based on orthogonal complement of a positive operator valued measure (POVM) on quantum phase. Then, we  construct quantum gates entanglers for different classes of multi-qubit states. In particular, we show that these operators can entangle multipartite state if they satisfy some conditions for W and GHZ classes of states. Finally, we explicitly give the W class and GHZ classes of quantum gate entanglers for four-qubit states.
\end{abstract}


\section{Introduction}
Multipartite entangled quantum states are very important resources in quantum information processing. A quantum computer can be constructed by quantum gates and one special class of these gates is called quantum gate entanglers. It also would be interesting if these quantum gate entanglers are able to produce any class of multipartite entangled quantum states. Recently, we have proposed different methods to construct quantum gate entanglers for multi-qubits states \cite{Hosh1}. We have also discussed some classes of quantum gate entangler based on different classes of positive operator on quantum phase (POVM) \cite{Hosh2}. In this paper, in continuation of previous works, we discuss the construction quantum gate entangler for multi-qubit states. In particular, in  this section,  we present necessary tools for designing some classes of quantum gate entanglers. Then, in section \ref{sec:1} we will  construct quantum gate entanglers for multi-qubit states. Finally, in section \ref{sec:2}, we will in detail discuss quantum gate entanglers for four-qubit states.

Let a state of a pure multi-qubit quantum system $\mathcal{Q}
=\mathcal{Q}_{1}\mathcal{Q}_{2}\cdots\mathcal{Q}_{m}$ be given by
\begin{equation}\ket{\Psi}=\sum^{1}_{x_{m},x_{m-1},\ldots,
x_{1}=0}\alpha_{x_{1}x_{2}\cdots x_{m}}\ket{x_{1}x_{2}\cdots
x_{m}},
\end{equation}
defined on a  Hilbert space $
\mathcal{H}_{\mathcal{Q}}=\mathcal{H}_{\mathcal{Q}_{1}}\otimes
\mathcal{H}_{\mathcal{Q}_{2}}\otimes\cdots\otimes\mathcal{H}_{\mathcal{Q}_{m}}
$. Moreover, let $Z=\left(
                                                        \begin{array}{cc}
                                                          1 & 0 \\
                                                          0 &-1 \\
                                                        \end{array}
                                                      \right)$ be
                                                      a gate that
                                                      acts on a qubit as
                                                      $Z\ket{0}=\ket{0}$ and $Z\ket{1}=-\ket{1}$.
                                                      Then, a
controlled phase gate is defined by $CZ=\frac{1}{2}(I_{2}\otimes I_{2}+I_{2}\otimes Z+Z\otimes
I_{2}-Z\otimes Z)$, where $I_{2}$ is a 2-by-2 identity matrix \cite{Ohmi,Chen}.

Next, we will review the basic definition and property of a general POVM on quantum phase
\cite{Hosh3}. This POVM is a set of linear operators
$\Delta(\varphi_{1,2},\ldots,\varphi_{1,N_{j}},\varphi_{2,3}$ $,\ldots,
\varphi_{N_{j}-1,N_{j}})$ that gives the probabilistic
measurement of a state $\rho_{Q_{j}}$ on the Hilbert space
$\mathcal{H}_{Q_{j}}$
\begin{eqnarray}&&p(\varphi_{1,2},\ldots,
\varphi_{1,N_{j}},\varphi_{2,3},\ldots,\varphi_{N_{j}-1,N_{j}})\\\nonumber&&=
\mathrm{Tr}(\rho\Delta
(\varphi_{1,2},\ldots,\varphi_{1,N_{j}},\varphi_{2,3},
\ldots,\varphi_{N_{j}-1,N_{j}})),
   \end{eqnarray}
where
$(\varphi_{1,2},\ldots,\varphi_{1,N_{j}},\varphi_{2,3},\ldots,\varphi_{N_{j}-1,N_{j}})$
are the outcomes of the measurement of the quantum  phase. This
POVM satisfies the following properties,
$\Delta(\varphi_{1,2},\ldots,\varphi_{1,N_{j}},\varphi_{2,3},\ldots,\varphi_{N_{j}-1,N_{j}})$
is self-adjoint, positive, and  normalized, that is
\begin{eqnarray}&&\overbrace{\int_{2\pi}\cdots
\int_{2\pi}}^{N_{j}(N_{j}-1)/2}d\varphi_{1,2}\cdots
d\varphi_{1,N_{j}}d\varphi_{2,3}\cdots
d\varphi_{N_{j}-1,N_{j}}\\\nonumber&&
    \Delta(\varphi_{1,2},\ldots,\varphi_{1,N_{j}},\varphi_{2,3},\ldots,\varphi_{N_{j}-1,N_{_{j}}})
    =\mathcal{I}_{N_{j}},
   \end{eqnarray}
   where
 the integral extends over any $2\pi$ intervals.
A general and symmetric POVM in a single $N_{j}$-dimensional
Hilbert space $\mathcal{H}_{\mathcal{Q}_{j}}$ is given by
\begin{eqnarray}
&&\Delta(\varphi_{k_j,l_j})=
   \left(%
\begin{array}{ccccc}
  1 &e^{i\varphi_{1,2}}  & \cdots
  &  e^{i\varphi_{1,N_{j}-1}} &e^{i\varphi_{1,N_{j}}}\\
 e^{-i\varphi_{1,2}} &  1 & \cdots
 &  e^{i\varphi_{2,N_{j}-1}} &e^{i\varphi_{2,N_{j}}}\\
  \vdots&  \vdots&\ddots &\vdots& \vdots\\
    e^{-i\varphi_{1,N_{j}-1}} & e^{-i\varphi_{2,N_{j}-1}} &\cdots&1&e^{i\varphi_{N_{j}-1,N_{j}}}\\
  e^{-i\varphi_{1,N_{j}}} & e^{-i\varphi_{2,N_{j}}} &\cdots&e^{-i\varphi_{N_{j}-1,N_{j}}}&1\\
\end{array}%
\right),
\end{eqnarray}
where $k_{j}<l_{j}$. The POVM is a
function of the $N_{j}(N_{j}-1)/2$ phases
$(\varphi_{1_j,2_j},\ldots,$ $\varphi_{1_j,N_j},\varphi_{2_j,3_j},\ldots,\varphi_{N_{j}-1,N_{j}})$.
It is also possible to define POVM for a multipartite system by
simply forming the tensor product
\begin{eqnarray}\label{POVM}\nonumber
\Delta_\mathcal{Q}(\varphi_{k_{1},l_{1}},\ldots,
\varphi_{k_{m},l_{m}})&=&
\Delta_{\mathcal{Q}_{1}}(\varphi_{k_{1},l_{1}})
\otimes\cdots
\otimes\Delta_{\mathcal{Q}_{m}}(\varphi_{k_{m},l_{m}}),
\\
\end{eqnarray}
where, e.g., $\varphi_{k_{1},l_{1}}$ is the set of
POVMs phase associated with subsystems $\mathcal{Q}_{1}$, for all
$k_{1},l_{1}=1,2,\ldots,N_{1}$, where we need only to consider
when $l_{1}>k_{1}$.

\section{Quantum gate entangler for multipartite  states} \label{sec:1}
In this section, first we will review the construction of concurrence classes for a pure multi-qubit states
\cite{Hosh4}. Then, we will propose quantum gate entanglers for multi-qubit states.
The
unique structure of our POVM enables us to distinguish different
classes of multipartite states. In the $m$-partite case, the  off-diagonal elements of
the matrix
\begin{eqnarray}\nonumber
\widetilde{\Delta}_\mathcal{Q}(\varphi_{k_{1},l_{1}},\ldots,
\varphi_{k_{m},l_{m}})&=&
\widetilde{\Delta}_{\mathcal{Q}_{1}}(\varphi_{k_{1},l_{1}})
\otimes\cdots
\otimes\widetilde{\Delta}_{\mathcal{Q}_{m}}(\varphi_{k_{m},l_{m}}),
\\
\end{eqnarray}
 have phases that are sum
or differences of phases originating from two and $m$ subsystems.
That is, in the later case the phases of
$\widetilde{\Delta}_\mathcal{Q}(\varphi_{k_{1},l_{1}},\ldots,
\varphi_{k_{m},l_{m}})$ take the form
$(\varphi_{k_{1},l_{1}}\pm\varphi_{k_{2},l_{2}}
\pm\ldots\pm\varphi_{k_{m},l_{m}})$. Thus, we  have proposed linear operators for  the $W^{m}$
class based on our POVM which are sum and difference of phases of
two subsystems, i.e.,
$(\varphi_{k_{r_{1}},l_{r_{1}}}
\pm\varphi_{k_{r_{2}},l_{r_{2}}})$. That is,
for the $W^{m}$ class we have
\begin{eqnarray}
 \widetilde{\Delta}^{
W^{m}}_{\mathcal{Q}_{r_{1},r_{2}}}
&=&\mathcal{I}_{2_{1}} \otimes\cdots
\otimes\widetilde{\Delta}_{\mathcal{Q}_{r_{1}}}
(\varphi^{\frac{\pi}{2}}_{k_{r_{1}},l_{r_{1}}})\\\nonumber&&
\otimes\cdots\otimes \widetilde{\Delta}_{\mathcal{Q}_{r_{2}}}
(\varphi^{\frac{\pi}{2}}_{k_{r_{2}},l_{r_{2}}})\otimes\cdots\otimes
\mathcal{I}_{2_{m}}.
\end{eqnarray}
 For the $GHZ^{m}$ class, we also have proposed  linear
operators based on our POVM which are sum and difference of phases
of $m$-subsystems, i.e.,
$(\varphi_{k_{r_{1}},l_{r_{1}}}
\pm\varphi_{k_{r_{2}},l_{r_{2}}}\pm
\ldots\pm\varphi_{k_{m},l_{m}})$. That is, for the
$GHZ^{m}$ class we have
\begin{eqnarray}
 \widetilde{\Delta}^{
GHZ^{m}}_{\mathcal{Q}_{r_{1},r_{2}}}
&=&\widetilde{\Delta}_{\mathcal{Q}_{1}}
(\varphi^{\pi}_{k_{1},l_{1}})\otimes\cdots
\otimes\widetilde{\Delta}_{\mathcal{Q}_{r_{1}}}
(\varphi^{\frac{\pi}{2}}_{k_{r_{1}},l_{r_{1}}})\\\nonumber&&
\otimes\cdots\otimes \widetilde{\Delta}_{\mathcal{Q}_{r_{2}}}
(\varphi^{\frac{\pi}{2}}_{k_{r_{2}},l_{r_{2}}})\otimes\cdots\otimes
\widetilde{\Delta}_{\mathcal{Q}_{m}}
(\varphi^{\pi}_{k_{m},l_{m}}).
\end{eqnarray}
where by choosing
$\varphi^{\pi}_{k_{j},l_{j}}=\pi$ for all
$k_{j}<l_{j}, ~j=1,2,\ldots,m$, we get an operator which has the
structure of Pauli operator $\sigma_{x}$ embedded in a
higher-dimensional Hilbert space and coincides with $\sigma_{x}$
for a single-qubit.

Moreover, we have proposed linear operators for the $GHZ^{m-1}$ class
of $m$-partite states based on our POVM which are sum and
difference of phases of $m-1$-subsystems, i.e.,
$(\varphi_{k_{r_{1}},l_{r_{1}}}
\pm\varphi_{k_{r_{2}},l_{r_{2}}}
\pm\ldots\varphi_{k_{m-1},l_{m-1}}\pm\varphi_{k_{m-1},l_{m-1}})$.
That is, for the $GHZ^{m-1}$ class we have

\begin{eqnarray}\nonumber
\widetilde{\Delta}^{
GHZ^{m-1}}_{\mathcal{Q}_{r_{1}r_{2},r_{3}}}
&=& \widetilde{\Delta}_{\mathcal{Q}_{r_{1}}}
(\varphi^{\frac{\pi}{2}}_{k_{r_{1}},l_{r_{1}}})
\otimes\widetilde{\Delta}_{\mathcal{Q}_{r_{2}}}
(\varphi^{\frac{\pi}{2}}_{k_{r_{2}},l_{r_{2}}})
\otimes\widetilde{\Delta}_{\mathcal{Q}_{r_{3}}}
(\varphi^{\pi}_{k_{r_{3}},l_{r_{3}}})
\otimes\cdots
\otimes\\&&\widetilde{\Delta}_{\mathcal{Q}_{m-1}}
(\varphi^{\pi}_{k_{r_{m-1}},l_{r_{m-1}}})\otimes\mathcal{I}_{2_{m}}
,
\end{eqnarray}
where $1\leq r_{1}<r_{2}<\cdots<r_{m-1}<m$.

Based on these operators we have constructed concurrence classes for multi-qubit states.
For example, for $W^{m}$  class  let
\begin{eqnarray}
   \mathcal{C}(\mathcal{Q}^{W^{m}}_{r_{1},r_{2}})&=&
    \sum_{\forall k_{1},l_{1},\ldots,k_{m},l_{m}}
    \left|\langle \Psi\ket{\widetilde{\Delta}^{
W^{m}}_{\mathcal{Q}_{r_{1},r_{2}}}\mathcal{C}_{m}\Psi}
\right|^{^{2}}.
\end{eqnarray}
Then the $W^{m}$ class concurrence is given by
\begin{eqnarray}
\mathcal{C}(\mathcal{Q}^{W^{m}}_{m})&=&
    \left(\mathcal{N}^{W}_{m}\sum^{m}_{r_{2}>r_{1}=1}\mathcal{C}(\mathcal{Q}^{W^{m}}_{r_{1},r_{2}}
    )\right)^{1/2},
\end{eqnarray}
where $\mathcal{N}^{W}_{m}$ is a normalization constant. Note that
for $m$-partite states the $W^{m}$ class concurrences are zero
only for completely separable states.

Now, we propose the following class of quantum gate entanglers for multi-qubit states.
Let $\mathcal{Z}_{2^{m}\times 2^{m}}$
be defined by
\begin{eqnarray}\label{Rmat}
\mathcal{Z}_{2^{m}\times2^{m}}&=&\mathcal{Z}^{0}_{2^{m}\times2^{m}}\mathcal{Z}^{1}_{2^{m}\times2^{m}}
\cdots\mathcal{Z}^{2^{m-1}-1}_{2^{m}\times2^{m}}\\\nonumber&=&\left\{
  \begin{array}{ll}
    \bigoplus^{2^{m-1}-1}_{x=0} U_{x}, & \hbox{if} ~~U_{x}=\left(
                                                 \begin{array}{cc}
                                                   \alpha_{x} & 0 \\
                                                   0 & \alpha_{x+1} \\
                                                 \end{array}
                                               \right) \\
    \bigoplus^{2^{m-1}-1}_{x=0}U_{x}S_{x}
, & \hbox{if}~~U_{x}=\left(
                                                 \begin{array}{cc}
                                                  0& \alpha_{x}  \\
                                                  \alpha_{x+1}&0 \\
                                                 \end{array}
                                               \right)
  \end{array}
\right.,
\end{eqnarray}
where $S_{x}=\left(
                                                 \begin{array}{cc}
                                                   0 & 1 \\
                                                   1& 0 \\
                                                 \end{array}
                                               \right)$ for all $x=x_{m}2^{m-1}+x_{m-1}2^{m-2}+\cdots
+x_{1}2^{0}$ is a swap gate.
Then,
$\mathcal{Z}_{2^{m}\times
2^{m}}H^{\otimes
m}\ket{0}^{\otimes m}$
 is entangled  and belong to $\mathrm{X}^{m}$ class if
elements of $\mathcal{Z}_{2^{m}\times 2^{m}}$ satisfy
$  \mathcal{C}(\mathcal{Q}^{X^{m}}_{r_{1},r_{2}})\neq0$, where e.g., $\mathrm{X}^{m}$ denotes the $\mathrm{W}^{m}$ or $\mathrm{GHZ}^{m}$ classes of multipartite systems. In the next section we will give an illustrative example of four-qubit quantum gate entanglers.

\section{Quantum gate entangler for four-qubit states
}\label{sec:2}For general four-partite states we have three
different joint phases in our POVM which give
 $W^{4}$, $GHZ^{3}$, and $GHZ^{4}$ class of operators.
 For the $W^{4}$ class, we have six types
of entanglement, so there are six  operators corresponding to
entanglement between $\mathcal{Q}_{1}\mathcal{Q}_{2}$,
$\mathcal{Q}_{1}\mathcal{Q}_{3}$,
$\mathcal{Q}_{1}\mathcal{Q}_{4}$,
$\mathcal{Q}_{2}\mathcal{Q}_{3}$,
$\mathcal{Q}_{2}\mathcal{Q}_{4}$, and
$\mathcal{Q}_{3}\mathcal{Q}_{4}$ subsystems. The linear operator
corresponding to $\mathcal{Q}_{1}\mathcal{Q}_{2}$ is given by
\begin{eqnarray}\nonumber
 \widetilde{\Delta}^{
W^{4}}_{\mathcal{Q}_{1,2}}&=&\widetilde{\Delta}_{\mathcal{Q}_{1}}
(\varphi^{\frac{\pi}{2}}_{k_{1},l_{1}})
\otimes\widetilde{\Delta}_{\mathcal{Q}_{2}}
(\varphi^{\frac{\pi}{2}}_{k_{2},l_{2}})
\otimes\mathcal{I}_{2_{3}}\otimes\mathcal{I}_{2_{4}}.
\end{eqnarray}
$
 \widetilde{\Delta}^{
W^{4}}_{\mathcal{Q}_{1,3}} $, $
 \widetilde{\Delta}^{
W^{4}}_{\mathcal{Q}_{1,4}}$, $
 \widetilde{\Delta}^{
W^{4}}_{\mathcal{Q}_{2,3}} $, $
 \widetilde{\Delta}^{
W^{4}}_{\mathcal{Q}_{2,4}} $, and $
 \widetilde{\Delta}^{
W^{4}}_{\mathcal{Q}_{3,4}}$
 are defined in similar way.
Now, for a pure four-qubit quantum system
let
\begin{eqnarray}
\mathcal{C}(\mathcal{Q}^{W^{4}}_{r_{1},r_{2}})&=&
\sum_{\forall k_{1},l_{1},\ldots,k_{4},l_{4}}
    \left|\langle \Psi\ket{\widetilde{\Delta}^{
W^{4}}_{\mathcal{Q}_{r_{1},r_{2}}}
\mathcal{C}_{4}\Psi}\right|^{^{2}},
\end{eqnarray}
Then, the state
$\mathcal{Z}_{2^{4}\times
2^{4}}H^{\otimes
4}\ket{0}^{\otimes 4}$
 is entangled  and belong to $\mathrm{W}^{4}$ class if
elements of
\begin{eqnarray}\label{Rmat}
\mathcal{Z}_{2^{4}\times2^{4}}&=&\mathcal{Z}^{0}_{2^{4}\times2^{4}}\mathcal{Z}^{1}_{2^{4}\times2^{4}}
\cdots\mathcal{Z}^{7}_{2^{4}\times2^{4}}\\\nonumber&=&\left\{
  \begin{array}{ll}
    \bigoplus^{7}_{x=0} U_{x}, & \hbox{if} ~~U_{x}=\left(
                                                 \begin{array}{cc}
                                                   \alpha_{x} & 0 \\
                                                   0 & \alpha_{x+1} \\
                                                 \end{array}
                                               \right) \\
    \bigoplus^{7}_{x=0}U_{x}S_{x}
, & \hbox{if}~~U_{x}=\left(
                                                 \begin{array}{cc}
                                                  0& \alpha_{x}  \\
                                                  \alpha_{x+1}&0 \\
                                                 \end{array}
                                               \right)
  \end{array}
\right.,
\end{eqnarray}
satisfy
$  \mathcal{C}(\mathcal{Q}^{W^{4}}_{r_{1},r_{2}})\neq0$.
The second class of four-partite state that we want to consider is
the $GHZ^{3}$ class. For this class, we have four types of
entanglement. These linear operators are given by
\begin{eqnarray}\nonumber
 \widetilde{\Delta}^{
GHZ^{3}}_{\mathcal{Q}_{12,3}}&=&
\widetilde{\Delta}_{\mathcal{Q}_{1}}
(\varphi^{\frac{\pi}{2}}_{k_{1},l_{1}})
\otimes\widetilde{\Delta}_{\mathcal{Q}_{2}}
(\varphi^{\frac{\pi}{2}}_{k_{2},l_{2}})
\otimes\widetilde{\Delta}_{\mathcal{Q}_{3}}(\varphi^{\pi}_{k_{3},l_{3}})
\otimes\mathcal{I}_{2_{4}},
\end{eqnarray}
\begin{eqnarray}\nonumber
 \widetilde{\Delta}^{
GHZ^{3}}_{\mathcal{Q}_{12,4}}&=&
\widetilde{\Delta}_{\mathcal{Q}_{1}}
(\varphi^{\frac{\pi}{2}}_{k_{1},l_{1}}) \otimes
\widetilde{\Delta}_{\mathcal{Q}_{2}}
(\varphi^{\frac{\pi}{2}}_{k_{2},l_{2}})
\otimes\nonumber\mathcal{I}_{2_{3}}
\otimes\widetilde{\Delta}_{\mathcal{Q}_{4}}(\varphi^{\pi}_{k_{4},l_{4}}),
\end{eqnarray}
$
 \widetilde{\Delta}^{
GHZ^{3}}_{\mathcal{Q}_{13,4}}$
 and
$ \widetilde{\Delta}^{
GHZ^{3}}_{\mathcal{Q}_{23,4}}
$ can be defined in similar way. Moreover, with
\begin{eqnarray}
   \mathcal{C}(\mathcal{Q}^{GHZ^{3}}_{r_{1}r_{2},r_{3}})&=&
    \sum_{\forall k_{1},l_{1},\ldots,k_{4},l_{4}}
    \left|\langle \Psi\ket{\widetilde{\Delta}^{
GHZ^{3}}_{\mathcal{Q}_{r_{1}r_{2},r_{3}}
}\mathcal{C}_{4}\Psi}\right|^{^{2}},
\end{eqnarray}
the state
$\mathcal{Z}_{2^{4}\times
2^{4}}H^{\otimes
4}\ket{0}^{\otimes 4}$
 is entangled  and belong to $\mathrm{GHZ}^{3}$ class if
elements of $\mathcal{Z}_{2^{4}\times 2^{4}}$ satisfy
$  \mathcal{C}(\mathcal{Q}^{GHZ^{3}}_{r_{1},r_{2}})\neq0$.
 Next we are going to consider the $GHZ^{4}$class
 concurrence for general four-partite states. For the $GHZ^{4}$ class, we have  six linear operators corresponding
to entanglement between these subsystems. For example, the linear operator
corresponding to
$(\mathcal{Q}_{1}\mathcal{Q}_{2})\mathcal{Q}_{3}\mathcal{Q}_{4}$
is given by
\begin{eqnarray}
 \widetilde{\Delta}^{
GHZ^{4}}_{\mathcal{Q}_{1,2}}&=&
\widetilde{\Delta}_{\mathcal{Q}_{1}}
(\varphi^{\frac{\pi}{2}}_{k_{1},l_{1}})
\otimes\widetilde{\Delta}_{\mathcal{Q}_{2}}
(\varphi^{\frac{\pi}{2}}_{k_{2},l_{2}})\otimes
\widetilde{\Delta}_{\mathcal{Q}_{3}}(\varphi^{\pi}_{k_{3},l_{3}})
\otimes\widetilde{\Delta}_{\mathcal{Q}_{4}}(\varphi^{\pi}_{\mathcal{Q}_{4};k_{4},l_{4}}).
\end{eqnarray}
$
 \widetilde{\Delta}^{
GHZ^{4}}_{\mathcal{Q}_{1,3}},~
 \widetilde{\Delta}^{
GHZ^{4}}_{\mathcal{Q}_{1,4}},~
 \widetilde{\Delta}^{
GHZ^{4}}_{\mathcal{Q}_{2,3}}, $
$\widetilde{\Delta}^{
GHZ^{4}}_{\mathcal{Q}_{2,4}}$, and $
 \widetilde{\Delta}^{
GHZ^{4}}_{\mathcal{Q}_{3,4}} $ are
defined in a similar way. Now, let
\begin{eqnarray}
   \mathcal{C}(\mathcal{Q}^{GHZ^{4}}_{r_{1},r_{2}})&=&
    \sum_{\forall k_{1}<l_{1},\ldots,k_{4}<l_{4}}
    \left|\langle \Psi\ket{\widetilde{\Delta}^{
GHZ^{4}}_{\mathcal{Q}_{r_{1},r_{2}}
}\mathcal{C}_{4}\Psi}\right|^{^{2}}.
\end{eqnarray}
Then, for $\mathrm{GHZ}^{4}$ class the state
$\mathcal{Z}_{2^{4}\times
2^{4}}H^{\otimes
4}\ket{0}^{\otimes 4}$
 is entangled  if
elements of $\mathcal{Z}_{2^{4}\times 2^{4}}$ satisfy
$  \mathcal{C}(\mathcal{Q}^{GHZ^{4}}_{r_{1},r_{2}})\neq0$.

\section{Conclusion}
In this paper, we have constructed a class of quantum gate entanglers for multi-qubit quantum states. Our construction was based on complement of POVM on quantum phase. By considering the quantum relative phase of  multi-qubit states we were able to propose a set of unitary operators that could entangle different classes of multi-qubit states. The result is interesting for the construction of distributed quantum computing and quantum control.

\begin{flushleft}
\textbf{Acknowledgments:}
 This work was supported by the Swedish Research Council (VR).
\end{flushleft}


\end{document}